# Miniaturized Computational Dispersion-Engineered Silicon Photonic Vernier Caliper Spectrometer

*Hao Deng[+], Tong Lin[+]\*, Liu Li, Yan Fan, Ziyang Xiong, Zelong Chen, Shihua Chen, Junpeng Lu\*, and Zhenhua Ni*

[+]These authors contributed equally

Hao Deng, Yan Fan, Ziyang Xiong

School of Electronic Science and Engineering, Southeast University, Nanjing 210096, China.

Tong Lin

School of Electronic Science and Engineering, Southeast University, Nanjing 210096, China.

Key Lab of Modern Optical Technologies of Education Ministry of China, Soochow University, Suzhou 215006, China.

E-mail: lintong@seu.edu.cn

Liu Li

School of Material Science and Engineering, Southeast University, Nanjing 211189, China.

Zelong Chen

School of Integrated Circuits, Southeast University, Nanjing 210008, China.

Shihua Chen

School of Physics and Key Laboratory of Quantum Materials and Devices of Ministry of Education, Southeast University, Nanjing 210023, China.

Junpeng Lu, Zhenhua Ni

School of Electronic Science and Engineering, Southeast University, Nanjing 210096, China.

School of Physics and Key Laboratory of Quantum Materials and Devices of Ministry of Education, Southeast University, Nanjing 210023, China.

E-mail: phyljp@seu.edu.cn

Funding: National Natural Science Foundation of China (62105061, 12374301, and 62225404); National Key Research and Development Program of China (2024YFA1210500).

**Abstract:** The development of miniaturized spectrometers for cost-effective mobile applications remains challenging, as small footprints fundamentally degrade bandwidth and resolution. Typically, achieving high resolution necessitates extended and sophisticated optical paths for spectral decorrelation. These restrict bandwidth both physically (through resonant wavelength periodicity constraints) and mathematically (due to resulting ill-conditioned large matrix factorizations). Here, we report a spectrometer using a computational dispersion-engineered silicon photonic Vernier caliper. This deterministic design enables periodicity-suppressed orthogonal measurements by nature, thus overcoming the bandwidth-resolution-footprint limit of current chip-scale spectrometers. Leveraging the dispersion-engineered Vernier subwavelength grating microrings and factorization-free matrix computation, a spectral resolution of 1.4 pm is achieved throughout a bandwidth of >160 nm with a footprint of <55×35 μm² in a single detection channel, establishing the highest bandwidth-to-resolution-to-footprint ratio (>57 μm$^{-2}$) demonstrated to date. Furthermore, broadband densely overlapped molecular absorption spectra of hydrogen cyanide are precisely measured, resolving 49 R- and P-branch lines with linewidths ranging from 15 to 86 pm which is fundamentally challenging for compressive sensing approaches. Our chip-scale spectrometer provides a new path toward precise and real-time multi-species spectral analysis and facilitates their commercialization.

**1. Introduction**

Spectroscopy is indispensable in numerous scientific and industrial fields,[1] playing a vital role in biomedicine, environmental monitoring, food quality control, and astronomy applications.[2] Traditional benchtop spectrometers, although well-established with high spectral resolution and broad bandwidth, often suffer from inherent drawbacks, including bulky sizes, prohibitive costs, and demanding optical alignment procedures.[3] Over the past decade, a growing demand for compact, cost-effective, yet high-performance spectrometers[4] is driven by the urgent need for portable and integrated optical sensing platforms capable of operating robustly in diverse, real-world environments. In response, photonic integrated circuits, particularly silicon photonics[5] built on silicon-on-insulator (SOI) platforms, have emerged as transformative technologies for on-chip spectrometers. Benefiting from complementary metal oxide semiconductor (CMOS) compatibility,[6] high refractive index contrast,[7] monolithic integration with modulator and photodetectors,[8] and co-integration with electronic ICs,[9] these silicon photonic spectrometers offer significant advantages in terms of cost-efficiency, miniaturization, and diverse functionalities across the short wavelength infrared (SWIR) range-where the overtone bands of various functional groups overlap.

Inspired by conventional benchtop spectrometers and leveraging silicon photonics, a variety of compact spectrometers with reduced footprint and weight have been developed primarily in two approaches.[4,10,11] The first is dispersive spectrometry which disperses or filters the spectra to different channels, employing diffraction-based components-such as arrayed waveguide gratings (AWGs),[12,13] and echelle diffraction gratings (EDGs),[14] or interference-based components like microring resonators (MRRs),[15] Fabry-Perot cavities,[16] and photonic crystal nanobeam cavities).[17] Despite significant efforts to achieve high performance chip-scale spectrometers, a basic contradiction persists among device footprint, spectral resolution, and bandwidth. Specifically, grating-based spectrometers enhance resolution by increasing waveguide length differences in AWGs[12,13] or enlarging the grating-to-output-waveguide distance in EDGs,[14] inevitably expanding the footprint to the millimeter range with a resolution of >0.2 nm. In contrast, interference-based resonators utilize cavity resonance effects to effectively extend the optical path length, enabling finer spectral resolution with substantially smaller dimensions (i.e., the full width at half maximum (FWHM) is 5 pm for a 1.47-mm-long MRR[15]). This enhancement results in a resonant wavelength periodicity, i.e., FSR, which is the inverse of the cavity perimeter,[18] limiting the spectrometers' bandwidth.[11] In order to overcome the limitations posed by the FSR, various cross-dispersion strategies for FSR expansion have been proposed, such as dual-mode dispersion[19] and FSR dispersion in one MRR,[20] mode splitting dispersion in dual MRRs,[21] and integration of multiple cavities.[15,22,23] Despite these advances, their bandwidths do not exceed 100 nm due to strong waveguide dispersion in compact footprints.[24] Such dispersion induces resonance deviations and coupling-strength variations,[21] while methods requiring multiple filter channels[10] further compromise device compactness.

The second category is computational spectrometers, specifically Fourier transform and compressive sensing types. Although advanced algorithms enable them to bypass bandwidth-resolution trade-offs,[10,25–29] persistent miniaturization challenges exist. Exemplifying this, Fourier transform spectrometers (FTS) derive spectral information by applying Fourier transform to the interferograms generated via waveguide path-length modulation. In principle, FTS resolution $\Delta\lambda$ is also physically constrained by the inverse proportionality to waveguide length ($\Delta\lambda \propto 1/L$),[30] while concurrent signal-to-noise ratio (SNR) degradation arises from propagation loss-induced signal attenuation.[31] A representative demonstration is a thin-film lithium niobate FTS (2.05×0.8 cm²) achieving 190 pm resolution over 340 nm bandwidth through 1.02 m-length equivalent waveguides.[27] On the other hand, compressive sensing spectrometers (CSSs) sample the entire incident spectrum in a N-dimensional spatiotemporal dataset,[32] reconstructing target wavelength points (M) via compressive sensing theory[33] and regression algorithms. This under-determined inverse problem (N≪M) necessitates two critical priors: signal sparsity and orthogonal measurement bases, with the M×N response matrix dimensionality fundamentally constrained by the target bandwidth-to-resolution ratio (i.e., channel number (M)).[34] Implementations encompass out-of-plane diffracted random speckles,[35–37] 32/64 stratified waveguide filter arrays,[38,39] 1D/2D Mach-Zehnder interferometer arrays,[40,41] multi-cavity systems,[28,32,42,43] etc. Notably, a miniaturized CSS (1.48×0.15 mm²) achieves a combined bandwidth-to-resolution ratio of 65,000 (520 nm/8 pm) with four separate superluminescent diodes.[28] Fundamentally, solving these underdetermined problems relies on ill-conditioned matrix inversion, which is highly susceptible to numerical instability from input noise and dimensionality effects. The noise-blurring effect exacerbates with increasing the target channel number, severely lowering SNR. Critically, current implementations require either extensive detector arrays or multi-heater systems with large thermal isolation gaps for building new orthogonal bases, inevitably expanding the footprint. Furthermore, CSS applicability remains intrinsically limited to sparse spectra;[10,33,44] reconstructing complex continuous spectra induces severe distortions and spurious peaks.[44] In-hardware pre-sparse method[44] has been proposed for general spectra but significantly increases system complexity and measurement time.

In this work, we introduce and experimentally demonstrate a hybrid approach that overcomes the bandwidth-resolution-footprint trade-off by synergizing widely tunable narrowband filtering with computational spectrometry as shown in Figure 1a. This continuous time-interleaving optical Vernier Caliper intrinsically serves as orthogonal bases for spectral reconstruction, thereby rendering the response matrix inversion a critically determined problem. The compact device size (55×35 μm²) is guaranteed by two cascaded trapezoidal subwavelength grating microring resonators (TSWG-MRRs). To mitigate strong dispersion[24] caused by this small footprint, we achieve near-uniform resonator responses by employing TSWG-engineered couplers inside them. Leveraging TSWG dispersion control,[45] a tunable Vernier optical Vernier filter[46] breaks the FSR limitation, exhibiting a 160-nm working window (>16 times larger than the MRR's FSR). In addition, the narrow and distinctive Vernier resonances are simultaneously tuned gapless through the synchronized thermal-optic tuning to support a single detection channel with a single photodetector. This distinctive scanning strategy provides adequate orthogonal bases for computational spectral reconstruction at each wavelength step, significantly simplifying the matrix computation complexity. By deploying two GPU-accelerated spectral reconstruction algorithms, we resolve fine spectral peaks separated by 1.4 pm over the 160-nm bandwidth—surpassing state-of-the-art capabilities—and identify 49 gas-phase molecular absorption lines with accuracy exceeding a commercial benchtop spectrometer. This hybrid approach establishes a new pathway toward chip-scale spectrometers delivering ultra-high resolution and broadband performance in a minimal form factor.

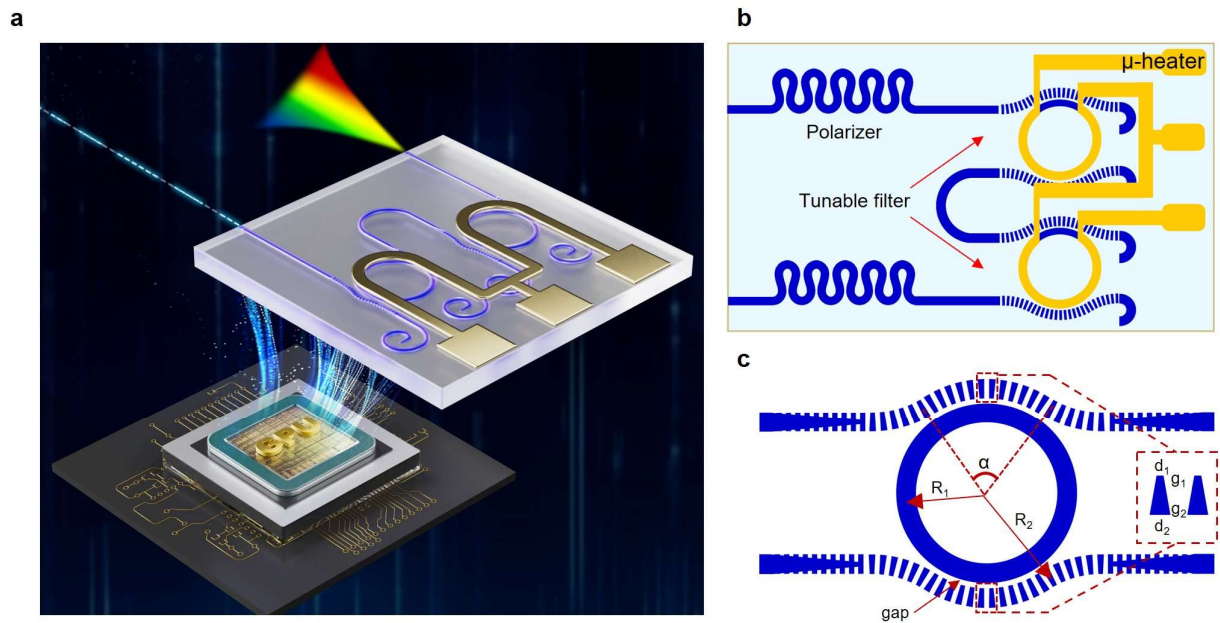

**Figure 1.** Illustration of the concept. a) The 3D artistic view of the proposed chip-scale spectrometer powered by a GPU circuitry. b) The schematic top view of the chip-scale spectrometer's photonic component layout. c) The structural diagram of one dispersion engineered trapezoid subwavelength grating microring resonator.

## 2. Principle and Device Design

The spectrometer's core photonic component—a wavelength-tunable Vernier filter—is implemented using two cascaded TSWG-MRRs with only 10-µm radii. Each resonator incorporates an integrated microheater atop for tuning, as schematically illustrated in Figure 1a,b. The physical properties of TSWG-MRRs directly determine the spectrometer's bandwidth, resolution, and footprint. Figure 1c details the schematic of a single TSWG-MRR, highlighting its pulley coupler configuration and critical parameters (e.g., $R_1$, $R_2$, α, gap, $d_1$, $d_2$, $g_1$, $g_2$). To broaden the spectrometer's working bandwidth using the Vernier effect, the two resonators incorporate curved TSWG waveguides and are meticulously designed with a slight radius offset. To balance the spectrometer's resolution and footprint, the resonators adopt circularly bent multi-mode waveguides with small radii to reduce scattering and bending losses. Therefore, a high Q-factor (i.e., a narrow FWHM) for an optimized FSR is attained. Compared to conventional MRRs, our resonator design utilizes curved TSWG waveguides that wrap around the circular multi-mode waveguides concentrically. These trapezoidal silicon pillars create an asymmetric effective index profile that minimizes the wavelength dependence. Therefore, these unique TSWG-assisted curved directional couplers (CDCs) inside the TSWG-MRR facilitate advanced dispersion engineering and enable nearly uniform coupling efficiency across a broad wavelength range. Additionally, integrated strip-to-SWG mode converters and TE-pass polarizers ensure exclusive transmission and analysis of TE-polarized light. We apply the 3D finite-difference time-domain (FDTD) simulation method and the particle swarm optimization algorithm for device optimization. The detailed structural parameters of two TSWG-MRRs are given in Supplementary Information S1.

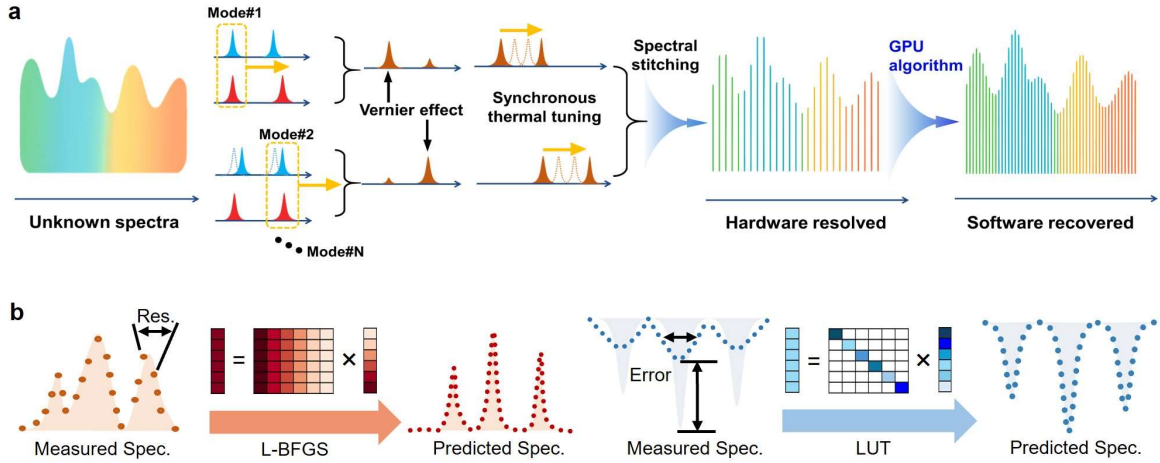

**Figure 2.** a) The working principle of the spectral reconstruction process includes Vernier resonance thermal tuning and software spectral recovery. b) Two types of spectral reconstruction algorithms with factorization-free methods: the peak deconvolution algorithm based on L-BFGS matrix inversion; the absorption-dip reconstruction algorithm based on the lookup table matrix transformer.

Figure 2a outlines the operational principle of the proposed spectrometer, combining the wavelength-tunable optical filtering with computational spectral reconstruction. An unknown spectrum is spectrally sliced into sub-bands defined by the FSR of the first TSWG-MRR. Each sub-band is then interrogated by a narrower resonance generated through the alignment of two

TSWG-MRR resonances under the Vernier effect.[46] This cascaded configuration achieves high-resolution filtering across a much broader bandwidth by extending the effective FSR to the least common multiple of individual resonators. By synchronously tuning them via integrated microheaters,[47] each sub-band is scanned by the tunable Vernier filter and a single photodetector captures the transmitted signal simultaneously. The entire spectrum is aggregated together from the converted electrical signals. Crucially, to overcome the inherent physical resolution limits imposed by the photonic components (TSWG-MRRs), GPU-accelerated spectral reconstruction algorithms de-convolve the system response by utilizing these orthogonal Vernier resonances, yielding a reconstructed spectrum with significantly enhanced resolution.

We propose two distinct spectral reconstruction algorithms tailored for wavemeter and absorbance spectroscopic scenarios, as shown in Figure 2b. The first is a peak deconvolution algorithm that improves our spectrometer's spectral resolution to the picometer scale, surpassing the fundamental resolution limits of its photonic components. This algorithm aims to reconstruct high-resolution multiple-peak spectra by iteratively optimizing spectral model parameters to minimize the mean squared error (MSE) between the measured and the predicted power spectra. The optimization employs the Limited-memory Broyden-Fletcher-Goldfarb-Shanno (L-BFGS) algorithm for its efficacy in solving nonlinear inverse problems inherent to spectral reconstruction tasks.[48] Formally, the target spectral intensity $S$ is reconstructed from the response matrix $H$ and the measured spectrum $O$ via: $S = H^{-1}O$. The response matrix $H$ that characterizes the system's full spectral response is constructed rigorously from its optical resonator response, which is described using an empirical Lorentzian function:

$$L(\lambda) = \frac{1}{1 + \left(\frac{\lambda - \lambda_0}{\gamma}\right)^2} \tag{1}$$

where $\lambda_0$ is the central wavelength of one resonance, and $\gamma$ represents the TSWG-MRR's FWHM. Each matrix element corresponds to the response intensity at each sampled $\lambda$, serving as an input for the optimization algorithm. Even though the response matrix H is also ill-conditioned due to the deterministic resonance overlap, its inversion can be implemented without matrix factorization. We thus formulate a regularized objective function to solve the critical-determined nonlinear least-squares:

$$\min_{S} \|O - P_{pred.}\|_2^2 + W_{\text{reg}} R_{\text{smooth}}(S) + W_{\text{sharp}} R_{\text{sharp}}(S) \tag{2}$$

here, $\|\cdot\|_2$ represents the $l_2$-norm; $P_{pred.}$ is the predicted power spectrum; $R_{\text{smooth}}(S)$ and $R_{\text{sharp}}(S)$ represent smoothness and sharpness regularization terms with two weighting coefficients ($W_{reg}$ and $W_{sharp}$), respectively. These regularization terms enforce that the reconstructed spectra maintain spectral fidelity—preserving physically realistic smoothness while suppressing high-frequency artifacts. To address system nonlinearities, we compute $P_{pred.}$ using a soft-max approximation:

$$P_{pred.} = \left(\text{softmax}(\alpha \cdot H \cdot S(\lambda))\right)^2 \quad (3)$$

where α is a scaling parameter that balances approximation-smoothness tradeoffs for different measures and $S(\lambda)$ is parameterized as an exponential function of optimization variables to ensure non-negativity. The target spectrum is deduced through iterative MSE minimization, with implementation details provided in Supplementary Information S3.

An absorption-dip reconstruction approach created especially for gas phase absorption spectroscopy is the second. This approach significantly mitigates spectral distortion when analyzing densely overlapping fine-structure spectra without sufficient spectrometer resolution. This algorithm restores high-resolution spectra from low-resolution measurements with an $O(n)$ time complexity by referencing a pre-calibrated lookup table (LUT). Although insufficient spectral resolution causes shallow envelope troughs of gas phase absorption lines, both dip positions and envelope shapes remain exploitable for reconstruction. These features originate from the convolution of the TSWG-MRR's optical response (Equation 1) with the analytical model of gas absorption lines. For experimental validation, we adopt the Lorentzian profile as the gas model, providing a computationally efficient approximation of the Voigt profile[49] while retaining physical interpretability. To leverage this convolutional correspondence, we explore exhaustive convolutions of Vernier lineshapes with the gas Lorentzian profiles and produce the LUT entries which are indexed by three parameters: the spectrometer resolution, the FWHM and dip depth of the convoluted peak. This transposed convolution mechanism operates through a configurable diagonal matrix for error compensation: uncorrected spectral bands retain unity-valued diagonal elements, while bands requiring correction apply amplitude-ratio scaling factors derived from deviations between raw and LUT-retrieved reconstructions with weighted smoothing coefficients eliminating transitional discontinuities. This allows for the real-time reconstruction of gas phase absorption spectra with unprecedented accuracy. Both aforementioned algorithms can be implemented into a portable GPU or FPGA platform, replacing conventional computers to optimize size, weight, and power qualities.

## 3. Experimental Results and Discussion

The proposed chip-scale spectrometer is fabricated on the standard silicon-on-insulator (SOI) platform, utilizing a 220 nm thick silicon device layer atop a 3 μm buried oxide from a standard Multi-Project Wafer foundry process (Applied Nanotools Inc.). Figure 3a provides optical microscope images of the actual device, clearly showing the integrated cascaded TSWG-MRRs with a footprint of less than 0.002 mm². A pair of TE-mode polarizers (more details are given in Supplementary Information S2) are adopted to filter the TM modes induced by inverse taper edge couplers and the associated integrated microheater (Pad I, Pad II, Pad III) are placed about 2 μm above both TSWG-MRRs for efficient thermal tuning.

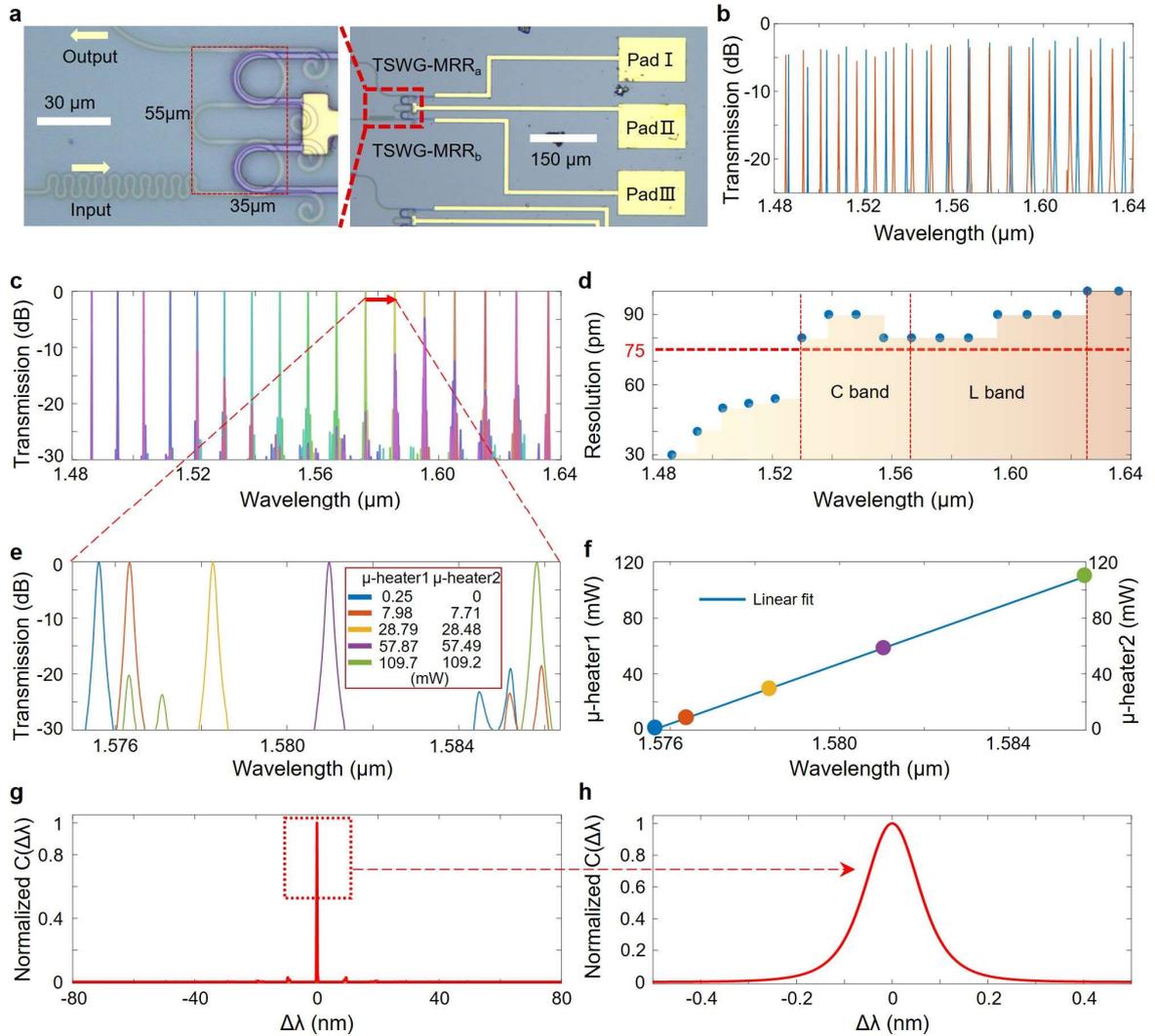

**Figure 3.** a) Optical microscope images of the fabricated silicon photonic spectrometer. The left one is the zoomed-in view of the cascaded SWG-MRRs and the TE-polarizer; the right one is the global view showing the three metal pads. b) The measured transmission spectra of the two SWG-MRRs (the red line denotes TSWG-MRR$_a$, the blue line denotes TSWG-MRR$_b$). c)

The normalized transmission spectra of the fabricated silicon photonic spectrometer's Vernier resonance that hops among different TSWG-MRR$_b$ FSRs. d) The measured silicon photonic spectrometer's physical resolution within the operating bandwidth from 1.48 μm to 1.64 μm (the average resolution is 75 pm). e) Five different transmission spectra of the tunable spectrometer's Vernier resonance within one TSWG-MRR$_b$ FSR tuning range near 1580 nm. f) The Vernier resonance wavelength changes linearly as the two microheater consumed powers vary. h) Calculated correlation functions [C(Δλ)] for the Vernier TSWG-MRRs around 1560 nm. i) Enlarged view of C(Δλ) around Δλ = 0.

We first characterize the individual and combined spectral responses of the cascaded resonators. Figure 3b displays the experimentally measured transmission spectra of the individual TSWG-MRR$_a$ and TSWG-MRR$_b$ using a tunable laser diode (Santec TSL-770), illustrating their distinct FSRs and high-quality factor resonances from 1480 nm to 1640 nm. The measured FSRs are about 8.53 nm and 9.22 nm respectively around 1550 nm. However, the FSRs change with the waveguide dispersion as the working bandwidth is very broad. Using the general formula proposed in[50], the calculated effective FSR is more than 160 nm which is >16 times larger than the TSWG-MRR$_b$'s FSR. Subsequently, we combine these responses to achieve the broadband Vernier spectrum by hopping across each FSR of TSWG-MRR$_b$. It's emphasized that only one Vernier resonance is toggled each time among the 160-nm-wide bandwidth as other resonances are suppressed because of the high side mode suppression ratio (SMSR). Meanwhile, achieving a narrower FWHM requires a smaller coupling coefficient, which inherently increases insertion loss at the drop port.[18] Therefore, the TSWG-MRRs' coupling coefficients need to be balanced between spectral resolution (FWHM) and insertion loss. Under our designed parameters, the total signal attenuation through the two TSWG-MRRs is measured at around 10 dB. The system link budget remains above the photodetector sensitivity threshold, ensuring reliable signal detection. Figure 3c shows the stitched experimental normalized Vernier resonance spectra, demonstrating uniform resonance peaks and high extinction ratios across the measurement range.

We quantify the spectrometer's physical resolution and validate the active tuning capabilities essential for rapid spectral acquisition. Figure 3d presents the fitted FWHM values of the Vernier resonances across the wavelength span, extending from the S-band into the U-band. We observe that the average resolution across the entire spectral range is 75 pm (indicated by the red dashed line), showcasing the device's inherent high spectral resolving power. The FWHM at the short wavelength range decreases from 80 pm to 30 pm, providing a surprisingly better resolution. The discrepancies between the simulation and experimental results are due to the increased losses caused by fabrication imperfections, which can be improved using a low index-contrast material platform.[28] To facilitate continuous spectral scan, we then demonstrate the ability to dynamically scan the Vernier resonances. Figures 3e illustrates five different tuning steps of a single Vernier resonance near 1576 nm with synchronized microheater powers and the SMSR maintains more than 20 dB. As shown in Figure 3f, the corresponding wavelength shift is linear with the applied microheater powers over one full FSR of TSWG-MRR$_b$, confirming precise and controllable thermal tuning. Furthermore, the microheater's response time is about 0.154 ms (more details are given in Supplementary Information S5), enabling high-speed spectral scanning at kHz rates as required for practical real-time applications. To quantify the correlation property of the wavelength channel, we calculate the correlation function [$C(\Delta\lambda)$] using the following equation:[21]

$$C(\Delta\lambda) = \frac{\langle I(\lambda)I(\lambda+\Delta\lambda)\rangle_\lambda}{\langle I(\lambda)\rangle_\lambda \langle I(\lambda+\Delta\lambda)\rangle_\lambda} - 1 \qquad (4)$$

where $\langle \cdot \rangle_\lambda$ denotes the average over $\lambda$ and $I(\lambda)$ is the intensity at the wavelength of $\lambda$. Figure 3g plots the normalized correlation of one Vernier resonance around 1560 nm across the 160 nm bandwidth: a single peak stands out while other modes are strongly suppressed due to the Vernier effect.[21] Its enlarged view is emphasized in Figure 3h, of which the linewidth corresponds to the FWHM.

We conduct two prevailing types of spectroscopic experiments to validate the high-resolution spectral reconstruction capabilities of our chip-scale spectrometer. Figure 4a illustrates the experimental setup employed for spectrum reconstruction. The unknown input spectrum is coupled into the spectrometer chip via one channel of a MFD conversion fiber array unit (FAU). The optical output from the chip is then directed to a photodetector (PD) and monitored in an oscilloscope. By programming the electrical powers applied to the microheaters on the chip using a 16-bit multi-channel programmable current source (SiliconExtream MCVS6400-A) and monitoring the PD (Thorlabs DET08CFC) output in the oscilloscope (Picoscope 5442D), the raw spectral data are acquired for sequential reconstruction offline with a personal laptop (Lenovo ThinkPad P16 Gen 2 AI 2024). This setup enables precise tuning of the Vernier resonances across the operational bandwidth, crucial for detailed spectral sampling.

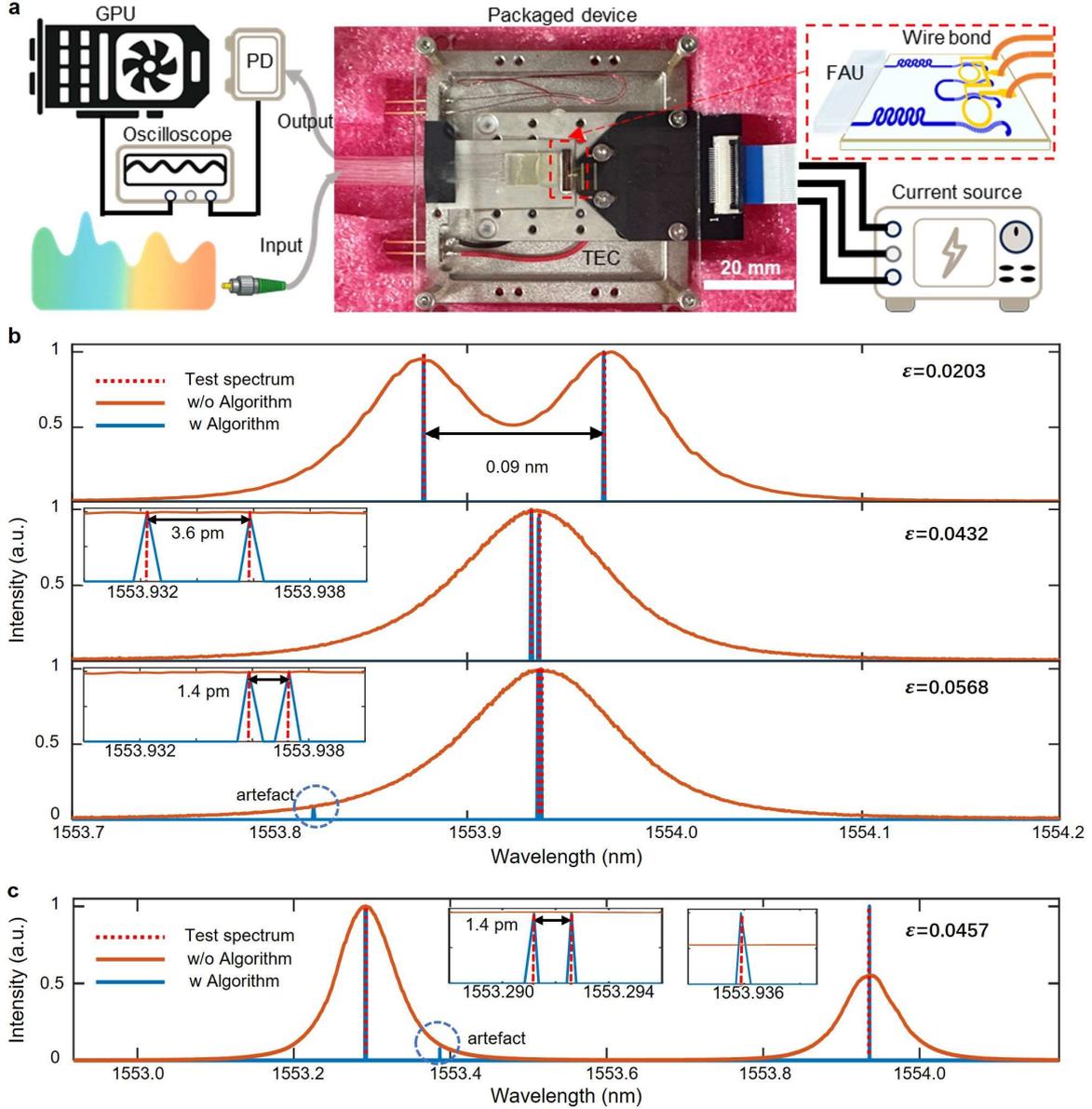

**Figure 4.** a) The experimental setup for measuring the unknown input spectra using the silicon photonic spectrometer chip in the temperature-controlled lab environment. The MFD conversion FAU is made using short sections of ultra-high-NA (UHNA) fibers that are spliced to pigtails of standard SM fibers, where the UHNA fiber's MFD matches that of the Si edge coupler. b) Retrieved spectra with three different double peak signals; their wavelength spacings are 90 pm, 3.6 pm, and 1.4 pm respectively. The red dotted line is the input source; the orang solid line is the physical reconstruction; the blue solid line is the algorithmically reconstruction. The normalized intensity of the artifact is 0.08. c) Retrieved spectra with triple peak signals; their wavelengths are 1553.2912 nm, 1553.2926 nm, and 1553.9359 nm. The normalized intensity of the artifact is 0.074. The red dotted line is the input source; the orang solid line is the physical reconstruction; the blue solid line is the algorithmically reconstruction.

With a spectrometer resolution of about 80 pm, we successfully resolve 1.4 pm spaced dual-peak and triple-peak signals using GPU-accelerated TSWG-MRRs. These signals were generated by coupling two or three separate tunable lasers operating at different wavelengths (their linewidths are less than hundreds of kHz) into fused fiber couplers, whose output was directly connected to our chip as the source to be measured. Figure 4b presents the spectrometer's reconstruction results for various wavelength separations between two narrow input peaks using both physically and algorithmically reconstructed methods. The top panel shows that the test spectrum with a wavelength spacing of 90 pm is clearly distinguished using either method. The linewidth of the retrieved spectrum is not as narrow as that of the laser sources, which is consistent with the device's intrinsic resolution (Figure 3d). Upon algorithm implementation, the spectrometer accomplishes a much narrower spectrum with the relative reconstruction error[32] $\varepsilon$ of 0.0203 as compared with the physical method; it resolves peak separations down to 3.6 pm (middle panel, inset) without presence of any unwanted artifact. The bottom panel further demonstrates resolution at 1.4 pm wavelength separation, where a tiny spectral artifact appears around 1553.821 nm. This artefact amplitude remains below 9% of the primary peak intensity. The $\varepsilon$ are 0.0432 and 0.0568 respectively. The consistency of the spectrometer's resolution across the entire bandwidth is verified rigorously at different wavelength regions[28,32] and more details are given in Supplementary Information S4. Further challenging the system, we conduct a triple-peak testing, wherein the spectral spacing among two closest peaks is also set at 1.4 pm. Figure 4c confirms the clear reconstruction of three laser peaks (1553.2912 nm, 1553.2926 nm, and 1553.9359 nm), indicating superior reconstruction accuracy with the $\varepsilon$ of 0.0457. A minor artifact appears around 1553.387 nm resembles with the dual-peak case, evidencing the noise origin hypothesis. This represents a 7-fold resolution enhancement over prior state-of-the-art reconstructive spectrometers, which achieve minimum dual-peak spacings of only 10 pm[41] and 400 pm[51] for triple-peak reconstruction, respectively.

To demonstrate the spectrometer's efficacy in practical applications, we successfully reconstruct broadband molecular absorption spectra of 49 P and R lines of gas-phase $H^{13}C^{14}N$ in the $2\nu_3$ band (1529–1565 nm, near-IR), as shown in Figure 5a. This task is often considered impossible for conventional reconstructive spectrometers which are only good for sparse spectra.[10,44] An amplified spontaneous emission source was fiber-coupled to the gas cell before entering into the chip, with subsequent measurements following the established experimental procedures. The top panel of Figure 5a provides the referenced spectrum acquired from another ultra-high-resolution tunable laser diode (Santec TSL570) with sub-MHz linewidth, the approach commonly known as the tunable diode laser absorption spectroscopy (TDLAS). The middle panel (light blue curve) shows a spectrum measured by a commercial optical spectrum analyzer (OSA, Anritsu MS9740B) at its highest resolution of 70 pm. Given that the linewidths of the 49 P and R lines range from 15 pm to 86 pm, the OSA's insufficient resolution results in significant spectral distortions. In stark contrast, our reconstructed results based on the algorithm-enhanced chip-scale system achieve sharper absorption features and deeper dips than the OSA and are on par with the TDLAS reference. Figure 5b quantifies the residual error per absorption line between our spectrometer and the OSA, defined as the absolute deviation in line depth from the TDLAS reference. Our system exhibits an average residual of 0.025—merely 24.9% of the OSA's error (0.1003). Detailed validation of the $H^{13}C^{14}N$ P2 line at 1543.809 nm (Figure 5c) confirms precise determination of both linewidth and peak position, which is critical for extracting gas parameters (e.g., concentration, pressure, temperature).[52] Cooperatively, these experimental results conclusively validate the co-design concept of photonic hardware and computational algorithms; this performance level is highly challenging to achieve with conventional reconstructive spectrometers.

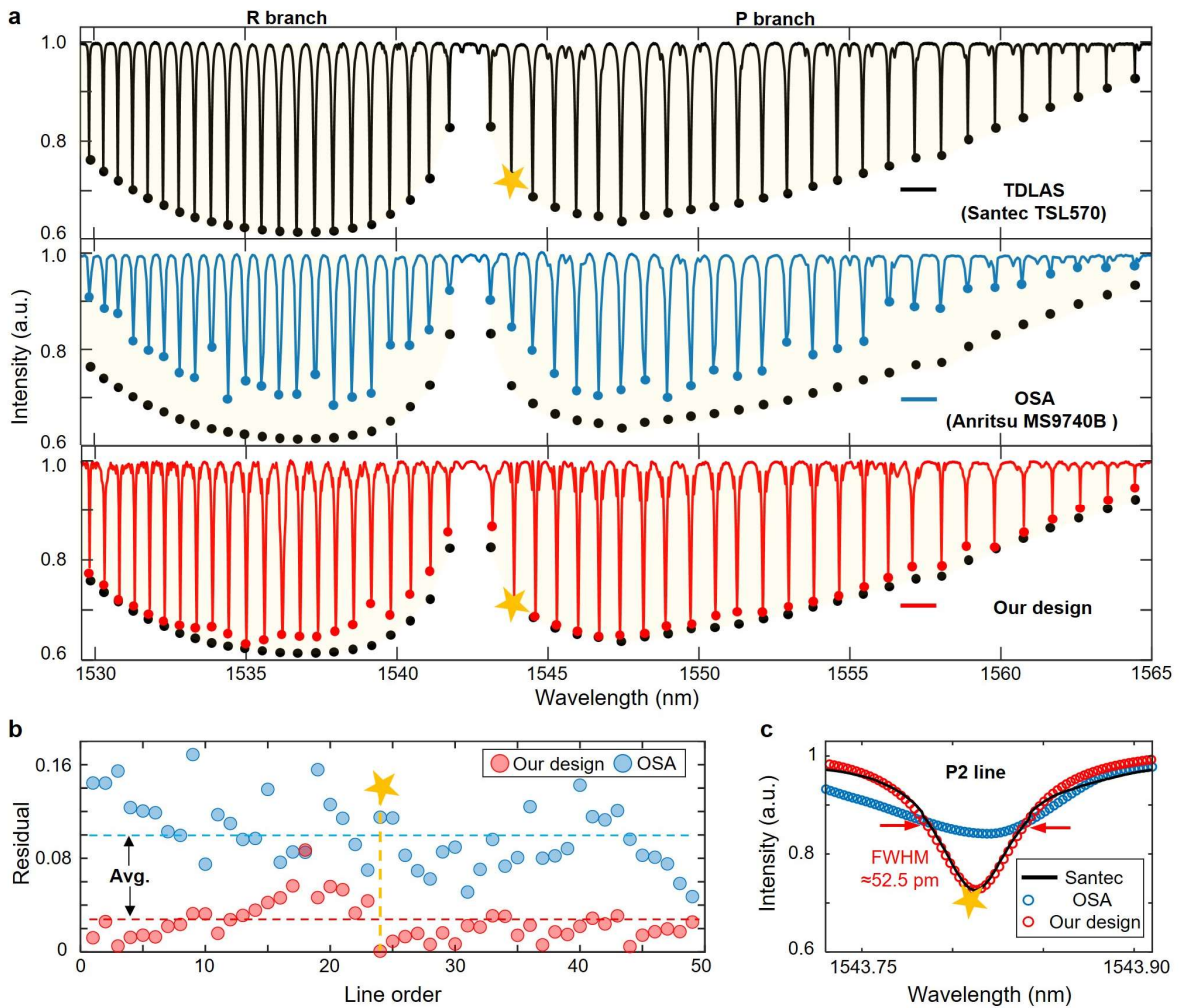

**Figure 5.** $H^{13}C^{14}N$ gas phase absorption spectroscopy experiment (The fiber coupled gas cell from Wavelength References has an absorption length of 16.5 cm and a pressure of 100 Torr). a) From the top to the bottom panel: the normalized measured spectra using the swept laser system (Santec TSL570 and MPM212) for reference, the commercial OSA (Anritsu MS9740B, 70 pm resolution), and our GPU algorithm-enhanced spectrometer. b) Residual comparison between our GPU-accelerated spectrometer (red points) and the OSA (light blue points). c) Zoomed in view of one reconstructed absorption line at 1543.809 nm ($H^{13}C^{14}N$ P2 line) with three different methods.

Figure 6a,b highlights state-of-the-art integrated spectrometers in terms of bandwidth ($BW$), resolution ($\delta\lambda$) and footprint ($s_f$). We define channel number (i.e., $N_{ch}=BW/\delta\lambda$) and channel-to-footprint ratio (i.e., CFR=$N_{ch}/\delta\lambda$) as systematic performance metrics.[21] Notably, dispersion-type spectrometers face a more pronounced three-way trade-off among these parameters than computational spectrometers. Our proposed co-design work, with an ultrasmall size of 55 × 35 μm² and a single spatial channel, is the first to simultaneously achieve $\delta\lambda$ < 1.5 pm, BW > 160 nm and CFR> 57.1 μm$^{-2}$. Figure 6a shows that one speckle enhanced FTS achieves the best resolution (140 MHz ≈ 1.12 pm), albeit with a limited 12-nm bandwidth.[26] It uses 64 cumbersome Mach-Zehnder Interferometers (MZIs) which is more than 5 mm long. Broader bandwidth requires even more MZIs, infeasible for chip-scale spectrometers. Regarding CFRs and $N_{ch}$ (Figure 6b), our work significantly outperforms competitors, as most CFRs are below 1 μm$^{-2}$. While the CSS[28] achieve greater bandwidth (520 nm) at 8 pm resolution, excelling in spectral reconstruction, their reliance on numerous microheaters for voltage-timing control substantially increases system size and power consumption, diminishing its CFR a lot. Another CSS propose a new path to enhance CFR based on compact disordered photonic molecules,[43] demonstrating a $N_{ch}$ of 12500 in a 70 × 50 μm² footprint and pushing the CFR to 3.57 μm$^{-2}$. Nevertheless, CSS reconstructive errors scale with spectral complexity, limiting reliable recovery to only particular and simple spectra via compressive sensing as discussed earlier.[44] Collectively, our co-design approach achieves superior performance across all three metrics. We further replicate reconstruction results based on a handheld development board (NVIDIA Jetson Nano B01), which are consistent with those using the 16-inch laptop. This confirms the feasibility for miniaturized spectral detection devices and details are presented in Supplementary Information S6.

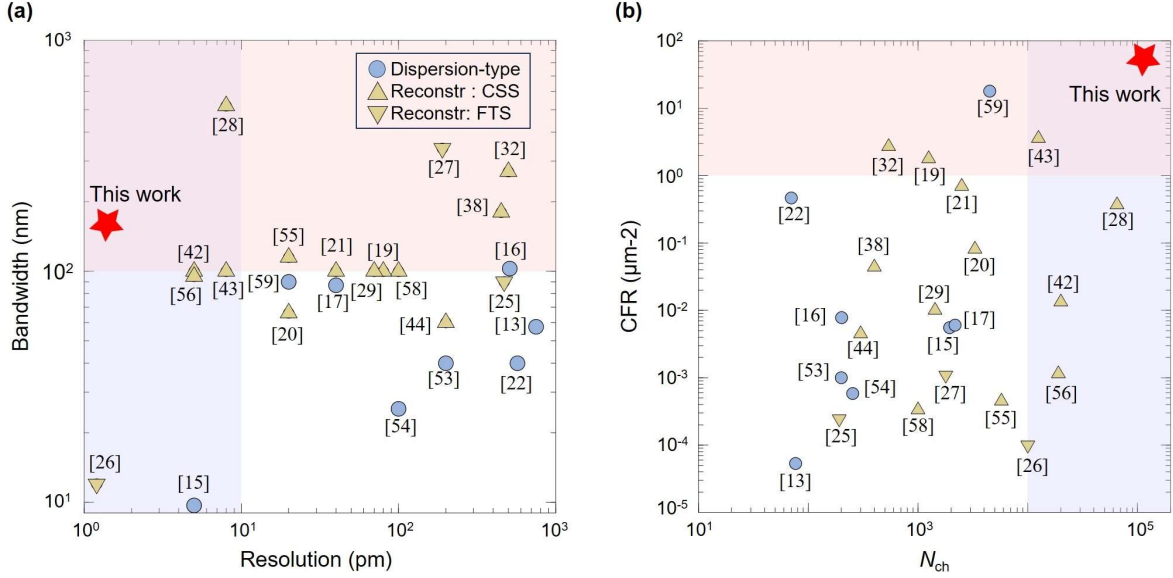

**Figure 6.** Performance comparison of various state-of-the-art chip-scale spectrometers with the focus on a) bandwidth versus resolution, b) channel-to-footprint ratios (CFR) versus the channel number ($N_{ch}$).

## 4. Conclusion

To summarize, we break the bandwidth-resolution-footprint trade-off in on-chip spectrometry through co-design of ultra-compact silicon photonic Vernier Caliper and GPU-accelerated computational algorithms. The integration of two thermally tunable TSWG-MRRs with TSWG-assisted CDCs allows for substantial performance improvements and simplified computation. The peak deconvolution algorithm enables the spectrometer to overcome its physical limitations and attain picometer-level resolution; the absorption-dip reconstruction algorithm facilitates precise measurements of 49 gas-phased $H^{13}C^{14}N$ absorption lines (R and P branches) with accuracy exceeding a commercial benchtop spectrometer. This hybrid approach offers an unprecedented bandwidth-to-resolution-to-footprint ratio (>57 μm$^{-2}$) demonstrated to date—a significant advance for integrated spectrometers. This performance can be easily enhanced by paralleling more devices for a much broader spectral range,[28] even extend from SWIR into Mid-IR. By overcoming the limitations of conventional dispersive and computational chip-scale spectrometers, this work paves the way for high-resolution broadband on-chip spectroscopy, maintaining compactness and cost-efficiency. The technology shows immediate promise for gas sensing, chemical synthesis, biological cell-sorting, and lab-on-a-chip.

**Acknowledgements**

This article was funded by the National Natural Science Foundation of China (62105061, 12374301, 62225404), the National Key Research and Development Program of China (2024YFA1210500), and the Key Lab of Modern Optical Technologies of Education, Ministry of China, Soochow University. The authors thank SJTU-Pinghu Institute of Intelligent Optoelectronics for their assistance with the chip packaging.

**Author contributions**

H.D. and T.L. conceived the idea, designed the device. H.D. performed numerical simulations, and developed the reconstruction algorithm. H.D., T.L., and Y.F. conducted the experiment. J.P.L and Z.H.N. supervised the project. H.D. and T.L. wrote the manuscript with support from all co-authors.

**Data Availability Statement**

All the data and methods needed to evaluate the conclusions of this work are presented in the main text and Supplementary Information. Additional data can be requested from the corresponding author.

**Conflict of interest**

The authors declare no competing interests.

**Supplementary Information**

The online version contains supplementary material is available.